\documentclass[12pt,a4paper]{article}
\usepackage[dvips]{epsfig}
\usepackage{multicol}
\usepackage{xspace}

\begin{document}


\begin{center}
{\Large \bf  Onset of Deconfinement and 
Critical Point - } 
\end{center}
\begin{center}
{\Large \bf 
- Future Ion Program at the CERN SPS\footnote{
invited talk given at NATO Advanced Research Workshop, 
September 10-16, 2005, Yalta, Ukrine}}
\end{center}

\vspace{1cm}
\begin{center}
{\large Marek Ga\'zdzicki} 
\end{center}
\begin{center}
{\large Institut f\"ur Kernphysik, University of Frankfurt, Germany and \\
Swietokrzyska Academy, Kielce, Poland}
\end{center}

\vspace{1cm}
\begin{abstract}
A new experimental program to study hadron production in collisions
of nuclei at the CERN SPS is presented. The program will focus on measurements
of fluctuations and long correlations with the aim to identify properties
of the onset of deconfinement and search for the critical point of
strongly interacting matter.
\end{abstract}

\section{Introduction}

Over the 50 years of CERN history numerous experimental
programs were devoted to a study of the properties of
strong interactions \cite{cern50}.
Rich data on collisions of hadron and ion beams were collected.
These results together with the data from other world
laboratories have significant impact on our understanding
of strong interactions  and at the same time lead to new
questions and define  directions of future experimental
and theoretical studies~\cite{villars}.
In particular, recently obtained exciting results 
\cite{afanasiev:2002mx,Gazdzicki:2004ef}
on nucleus-nucleus
collisions from the CERN SPS and the BNL RHIC suggest that the onset
of deconfinement is located at the low SPS energies \cite{Gazdzicki:1998vd}.
The most important open issues related to this finding are as follows.
Is it possible to
observe the predicted signals of the onset of deconfinement
in fluctuations and anisotropic flow measurements?
What is the nature of the transition from the anomalous energy
dependence seen in the central Pb+Pb collisions at the SPS energies
to the smooth dependence observed in p+p interactions?
Does the critical point of strongly interacting matter exist
and, if it does, where is it located?

In order to answer these questions  it was
proposed \cite{eoi} to perform a sequence of
new measurements of hadron production in collisions
of protons and nuclei with nuclear targets at CERN
by use of the upgraded NA49 apparatus \cite{na49_nim}.

\vspace*{0.2cm}
The NA49 apparatus at the  CERN~SPS served during the last 10 years
as a very reliable, large acceptance hadron spectrometer and
delivered unique high precision experimental data over the full
range of SPS beams (from proton to lead) \cite{na49_beam,Alt:2005zq}
and energies (from
20$A$ GeV to 200$A$ GeV) \cite{afanasiev:2002mx,Gazdzicki:2004ef}.
The most efficient and cost effective way to reach
the physics goals of the proposed new experimental program
is to
use  the upgraded NA49 detector, its reconstruction software
and over many years accumulated experience in physics analysis.

The paper is organized as follows. The key physics questions 
addressed by the new program are formulated in Section 2.
The basic requirements concerning the future measurements
are discussed in Section 3.
In Section 4 different experimental opportunities to
study nucleus-nucleus collisions are briefly reviewed.
The paper is closed by the summary, Section 5.

\section{Key Questions}

One of the key issues of contemporary physics is the understanding
of strong interactions and in particular the study of the
properties of strongly interacting matter in equilibrium.
What are the phases of this matter
and how do the transitions between them look like are questions
which motivate
a broad experimental and theoretical effort.
The study of high energy nucleus-nucleus collisions
gives us a unique possibility to address these questions
in well-controlled laboratory experiments.

\subsection*{Onset of Deconfinement}
\begin{figure}
\centerline{\includegraphics[width=9cm]{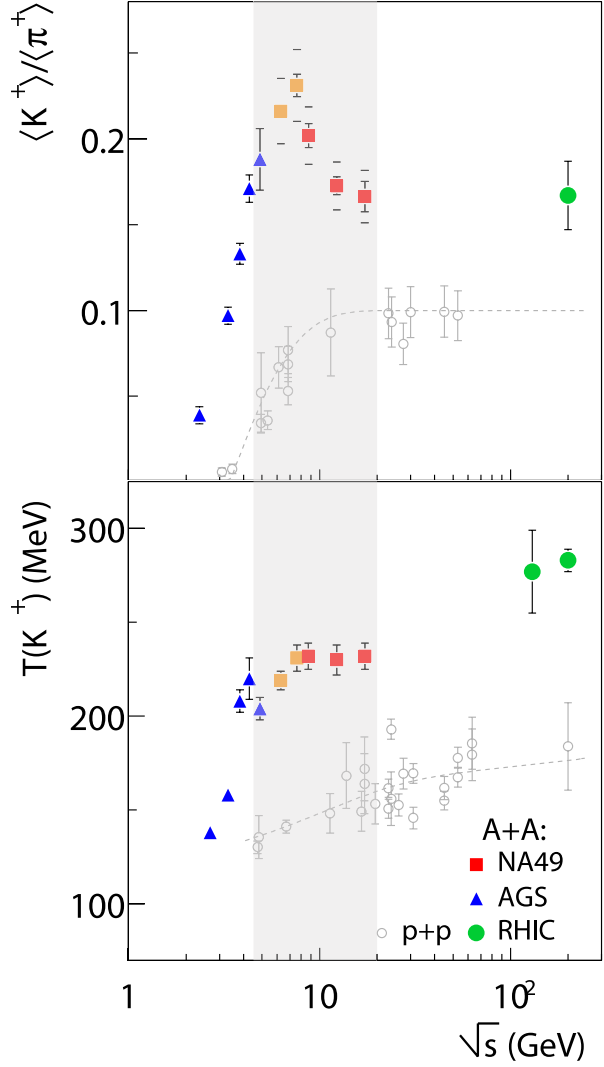}}
\caption[]{
Collision energy dependence
of the $K^+$ to $\pi^+$ ratio and the inverse slope
parameter of the transverse mass spectra
measured in central Pb+Pb and Au+Au collisions (solid symbols)
compared to results from p+p reactions (open dots). The changes
in the SPS energy range (solid squares) suggest the onset of
the deconfinement phase transition.
The energy region covered by the future measurements at the
CERN SPS is indicated by the gray band.
}%
\label{edep_cern_cour}
\end{figure}

Recent results 
\cite{afanasiev:2002mx,Gazdzicki:2004ef}
on the energy dependence of hadron production in central
Pb+Pb collisions at 20$A$, 30$A$, 40$A$, 80$A$ and 158$A$~GeV
coming from the energy scan program at the
CERN SPS serve as evidence for the existence
of a transition
to a new form of strongly interacting matter, the Quark Gluon
Plasma (QGP) in nature \cite{Gazdzicki:1998vd}.
Thus they are in agreement with the conjectures that at the top
SPS and RHIC energies the matter created at the early stage
of central Pb+Pb (Au+Au) collisions is in the state of
QGP~\cite{qgp_sps,qgp_rhic}.
The key results are summarized
in Fig.~\ref{edep_cern_cour}.

The most dramatic effect can be seen in the energy dependence of the ratio
$\langle K^+ \rangle/\langle \pi^+ \rangle$ of the mean multiplicities
of $K^+$ and $\pi^+$ produced per event in central
Pb+Pb collisions, which is plotted in the top panel of the figure.
Following a fast threshold rise, the ratio passes through a sharp
maximum in the SPS range and then seems to settle to a lower
plateau value at higher energies.
Kaons are the lightest strange hadrons and
 $\langle K^+ \rangle$ is equal to  about half of the number of all
anti-strange quarks
produced in the collisions.
Thus, the relative strangeness content of the
produced matter passes through a sharp maximum at the SPS
in nucleus-nucleus collisions. This feature is not observed
for proton-proton reactions.

A second important result is the constant value
of the apparent temperature  of $K^+$ mesons
in central Pb+Pb collisions
at low SPS energies
as shown in the bottom panel of the figure.
The plateau at the SPS energies is proceeded by a steep rise
of the apparent temperature
at the AGS and followed by a further increase
indicated by the RHIC data.
Very different behaviour is measured in proton-proton
interactions.

Presently, the sharp maximum and the following plateau in the energy dependence
of the $\langle K^+ \rangle/\langle \pi^+ \rangle$ ratio has only been
reproduced by the statistical model of the early stage
\cite{Gazdzicki:1998vd}
in which a first order phase transition is assumed.
In this
model the maximum
reflects the decrease in the number ratio of strange to non-strange
degrees of freedom and changes in their masses
when deconfinement sets in. Moreover, the
observed steepening of the increase in pion production is consistent
with the expected excitation of the quark and gluon degrees of freedom.
Finally, in the picture of the expanding fireball, the apparent temperature
is related to
the thermal  motion of the particles and their collective expansion velocity.
Collective expansion effects are expected to be important
only in heavy ion collisions
as they result from the pressure generated in the dense interacting matter.
The stationary value of the apparent temperature  of $K^+$ mesons
may thus indicate an approximate constancy of the early stage
temperature and  pressure in the SPS energy range
due to the coexistence of hadronic and deconfined phases,as in the case of the first order phase transition \cite{Gorenstein:2003cu,hama}.

Thus, the anomalies in the energy dependence of  hadron production
in central Pb+Pb collisions at the low SPS energies
serve as evidence for the
onset of deconfinement and the existence of QGP in nature.
They are consistent with the hypotheses that the observed transition is
of the first order.
The anomalies are not observed in p+p interactions and they are
not reproduced within hadronic models~\cite{hadronic_models}.

These results and their interpretation raise  questions
which can be answered only by new measurements.
Two most important open problems are:
\begin{itemize}
\item
{\bf is it possible to observe the predicted signals of the onset of
deconfinement in fluctuations \cite{mg_fluct}
and anisotropic flow \cite{Kolb:2000sd}?}
\item
{\bf what is the nature of the transition from the anomalous energy dependence
measured in central Pb+Pb collisions at SPS energies to the smooth dependence
measured in p+p interactions?}
\end{itemize}

\subsection*{Critical Point}

In the letter of Rajagopal, Shuryak, Stephanov and Wilczek
addressed to the SPS Committee one reads:
{\it ...  Recent theoretical developments suggest that a key
qualitative feature, namely a critical point 
(of strongly interacting matter) which in sense defines the
landscape to be mapped, may be within reach of discovery and analysis
by the SPS, if data is taken at several different energies.
The discovery of the critical point would in stroke transform the map
of the QCD phase diagram which we sketch below from one based only on 
reasonable inference from universality, lattice gauge theory and models
into one within a solid experimental basis. ...}
More detailed argumentation is presented below.

\begin{figure}[!ht]
\centerline{\includegraphics[width=12cm]{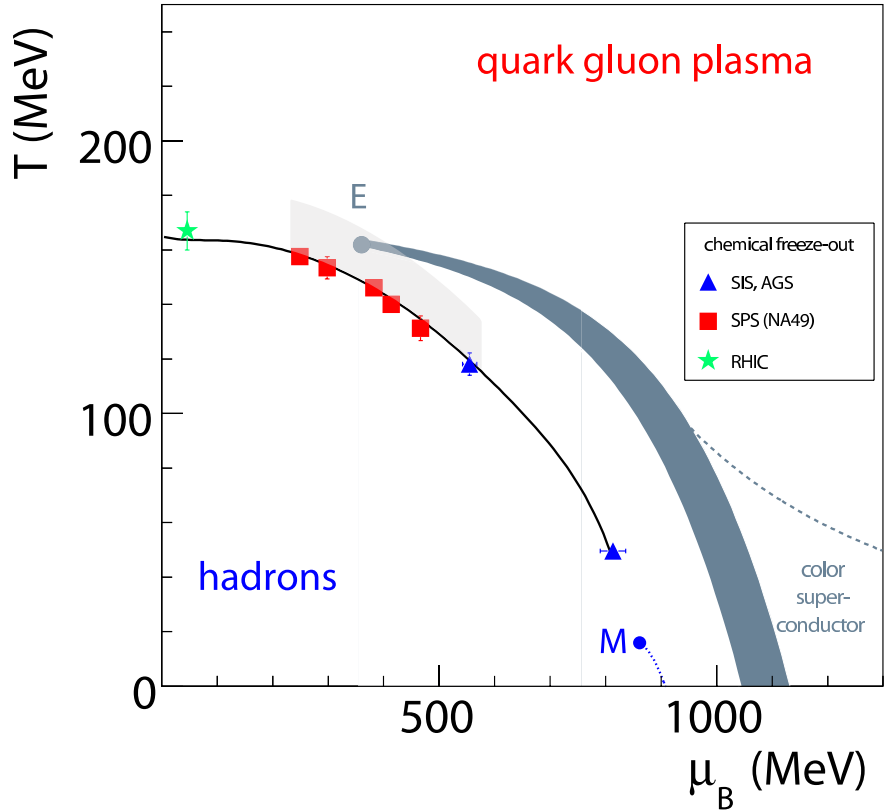}}
\caption[]{
The hypothetical phase diagram of strongly interacting
matter in the plane temperature, $T$, and baryonic chemical potential,
$\mu_B$. The end point {\bf E} of the first order transition strip is
the critical point of the second order.
The chemical freeze-out points extracted from the analysis
of hadron yields in central Pb+Pb (Au+Au) collisions at
different energies are plotted by the solid symbols.
The region covered by the future measurements at the CERN SPS
is indicated by the gray band.
}%
\label{phase}
\end{figure}

\vspace*{0.2cm}
Rich systematics of hadron multiplicities produced in nuclear collisions
can be reasonably well described by hadron gas models
\cite{Cleymans:1999cb,Braun-Munzinger:2003zd,becat}.
Among the model parameters fitted to the data are temperature, $T$, and baryonic
chemical
potential, $\mu_B$, of the matter at the stage of the freeze-out of hadron
composition (the chemical freeze-out). These parameters extracted
for central Pb+Pb collisions at the CERN SPS energies are plotted in
Fig. \ref{phase} together with the corresponding results for
higher (RHIC) and lower (AGS, SIS) energies. With increasing collision energy
$T$ increases and $\mu_B$ decreases.
A rapid increase of temperature is observed up to mid SPS energies, from the
top SPS energy ($\sqrt{s_{NN}}$ = 17.2 GeV) to the top RHIC energy
($\sqrt{s_{NN}}$ = 200 GeV) the temperature increases only by about 10 MeV.

\vspace*{0.2cm}
The sketch of the phase diagram of strongly interacting matter
in the ($T-\mu_B$) plane
as suggested by QCD-based considerations
\cite{Rajagopal:2000wf, Stephanov:2004wx} is
also shown in Fig.~\ref{phase}.
To a large extent these predictions are qualitative, as
QCD phenomenology at finite temperature and baryon number
is one of the least explored domains of the theory.
More quantitative results come from
lattice QCD calculations which can be performed
at $\mu_B = 0$.
They strongly suggest a rapid
crossover from the hadron gas to the QGP at the temperature
$T_C = 170-190$ MeV \cite{Karsch:2004wd,katz}, which seems to be somewhat higher
than the
chemical freeze-out temperatures of
central Pb+Pb collisions ($T =150-170$ MeV) \cite{jakko}
at the top SPS and RHIC energies.
The nature of the transition to QGP is expected
to change due to the  increasing
baryo-chemical potential.
At high potential the transition may  be of the
first order, where the end point of the first order transition
domain, marked $E$ in Fig.~\ref{phase},
is the critical point of the second order.
Recently even richer structure of the phase transition to QGP
was discussed within a statistical model of quark gluon bags
\cite{Gorenstein:2005rc}.
It was suggested that the line of the first order phase transition
at high $\mu_B$ is followed by the line of the second order
phase transition at intermediate $\mu_B$, and then by the lines of
''higher order transitions'' at low $\mu_B$.
A characteristic property of the second order phase transition
(the critical point or line) is
divergence of the susceptibilities.
An important test for a second-order phase transition
at the critical point is the validity 
of appropriate power laws in measurable quantities related to critical fluctuations.
Techniques associated with such measurements in nuclear collisions
have been developed recently \cite{antoniou:2005}
with emphasis on the sector of isoscalar di-pions ($\sigma$-mode)
as required by the QCD conjecture for the critical end point
in quark matter \cite{Rajagopal:2000wf}.
Employing such techniques in a study of nuclear collisions
at different energies at the SPS and with nuclei of different sizes,
the experiment may be confronted not only with the existence
and location of the critical point but also with the size of critical
fluctuations as given by the critical exponents of the QCD conjecture.

Thus when scanning the phase diagram a maximum
of fluctuations located in a domain close to the critical
point
($\Delta T \approx 15$ MeV and
$\Delta \mu_B \approx 50$ MeV \cite{Hatta:2002sj}) or the
critical line
should signal the second order phase transition.
The position of the critical region is uncertain,
but the best theoretical estimates based on lattice
QCD calculations locate it at $T \approx 160$ MeV and
$\mu_B \approx 360$ MeV
\cite{Fodor:2004nz,Allton:2005gk} as indicated in Fig. \ref{phase}.
It is thus in the vicinity  of the chemical freeze-out points
of central Pb+Pb collisions at the CERN SPS energies.

\begin{figure}[!ht]
\centerline{\includegraphics[width=11cm]{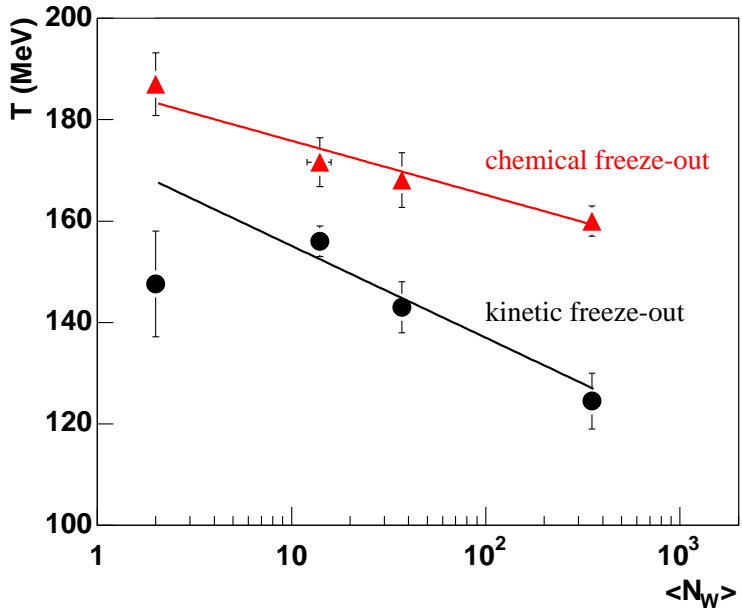}}
\caption[]{
The dependence of the chemical and kinetic temperatures
on the mean number of wounded nucleons for p+p, C+C, Si+Si and
Pb+Pb collisions at 158$A$ GeV.
}%
\label{temp_log}
\end{figure}

Pilot data \cite{na49_beam} on interactions of light nuclei (Si+Si, C+C and p+p)
taken by NA49 at 40$A$ and 158$A$ GeV indicate that the freeze-out
temperature increases with decreasing mass number, $A$, of the
colliding nuclei, see Fig.~\ref{temp_log}.
This means that the scan in the collision energy and
mass of the colliding nuclei allows us to scan  the
($T-\mu_B$) plane
\cite{Stephanov:1999zu}.

The experimental search for the critical point by investigating
nuclear collisions is justified at energies higher
than the energy of the onset of deconfinement.
This is because the energy density at the early stage of
the collision, which is relevant for the onset of deconfinement
is higher than the energy density at the freeze-out, which
is of the importance for the search for the critical point.
The only anomalies possibly related to the onset of deconfinement
are measured at $\sqrt{s_{NN}} \approx 8$ GeV
(see Fig.~\ref{edep_cern_cour}). This limits
a search for the critical point to an energy range
$\sqrt{s_{NN}} > 8$ GeV.
Fortunately, as discussed above and illustrated in Fig.~\ref{phase},
the best theoretical predictions locate the  critical point
in the ($T-\mu_B$)  region
accessible in nuclear collisions in this energy range.
Thus the new measurements at the CERN SPS can answer the
fundamental question:
\begin{itemize}
\item
{\bf does the critical point of strongly interacting matter
exist in nature and, if it does, where is it located?}
\end{itemize}

\section{General requirements}

The physics goals of the new experimental program with
nuclear beams at the CERN SPS presented in the previous
section require the energy scan in the whole SPS energy
range (10$A$-200$A$ GeV) with light and intermediate mass
nuclei. The measurements should be focused on the
precise study of fluctuations and anisotropic flow.
The first NA49 results on these subjects \cite{na49_flow,na49_pt,
na49_n,soft_balls,na49_k} suggest, in fact,
presence of interesting effects for collisions with
moderate number of participant nucleons and/or at low
collision energies.
However, a very limited set of data and serious resolution and
acceptance limitations do not allow firm conclusions.
The general physics needs when confronted with the
NA49 results and experience suggest improvements of the
current performance of the NA49 apparatus.
\begin{enumerate}
\item
The event collection rate should be significantly increased
in order to allow a fast registration of a sufficient statistics
for a large number of different reactions ($A$, $\sqrt{s_{NN}}$).
\item
The resolution in the event centrality determination based
on the measurement of the energy of projectile spectator
nucleons should be improved.
This is important for  high precision measurements of the
event-by-event fluctuations.
\item
The acceptance of the measurements of charged hadrons has to
be increased and made as uniform as possible.
\end{enumerate}

With this improved set-up we intend to register
C+C, Si+Si and In+In collisions at 10$A$,
20$A$, 30$A$, 40$A$, 80$A$, 158$A$ GeV and a typical
number of recorded events per reaction is $2 \cdot 10^6$.

\section{Experimental landscape}

\begin{figure}[!ht]
\centerline{\includegraphics[width=17cm]{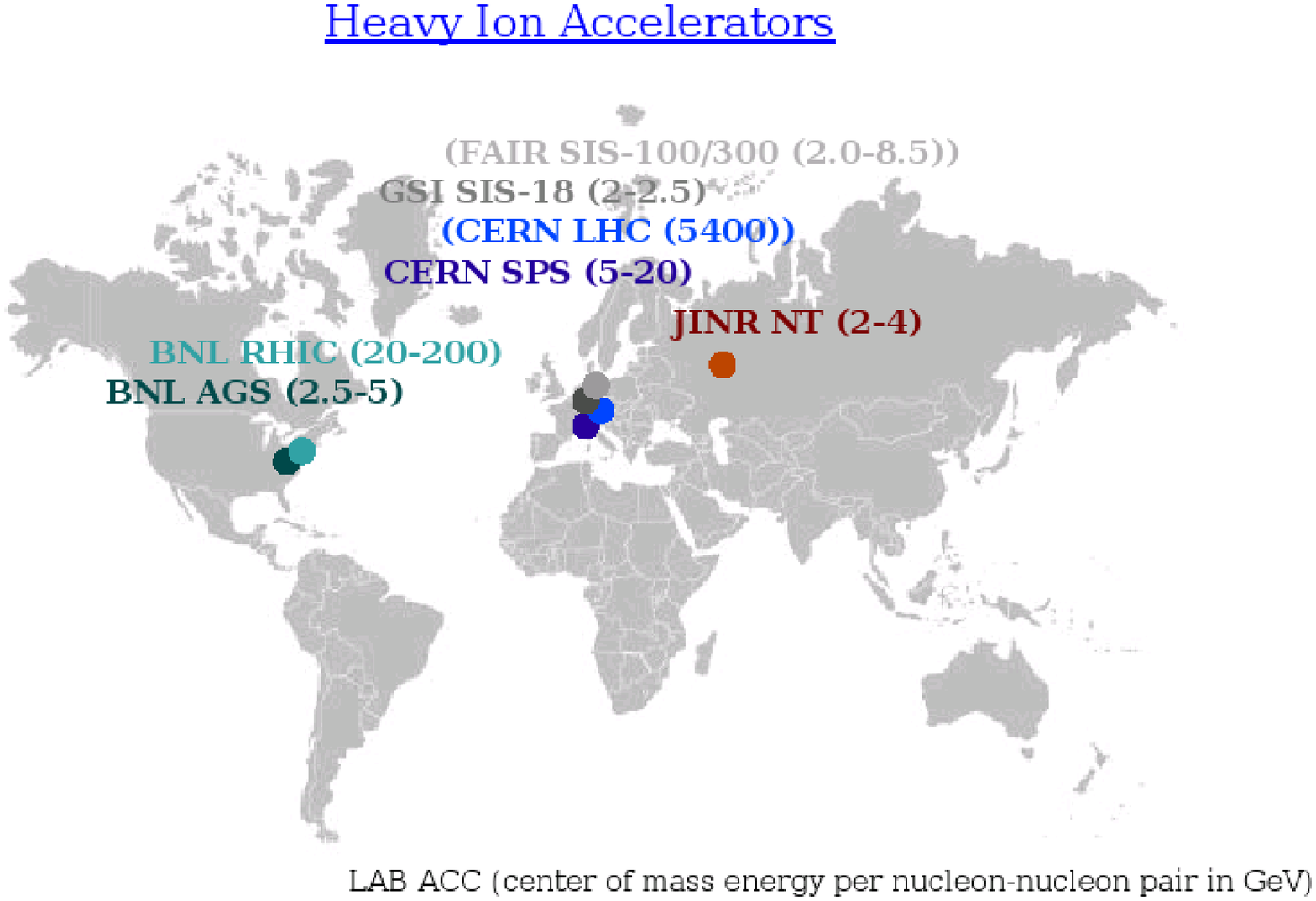}}
\caption[]{
The present and future accelerators of relativistic
nuclear beams. For each facility the name of the laboratory,
the name of the accelerator and  the energy
range (center of mass energy per nucleon-nucleon pair in GeV)
are given.
}%
\label{landscape}
\end{figure}

\begin{figure}[!ht]
\centerline{\includegraphics[width=15cm]{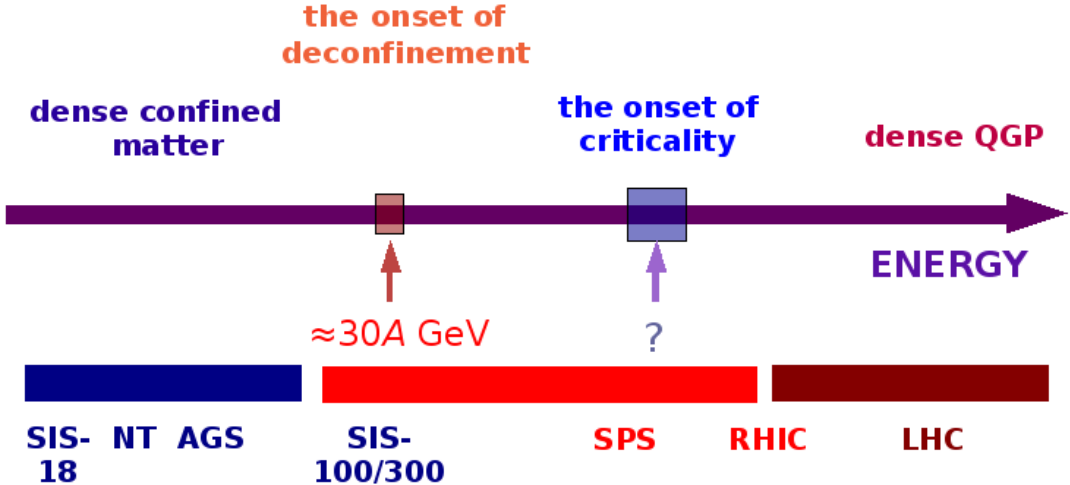}}
\caption[]{
The accelerators of relativistic nuclear beams
ordered according to the top collision energy and the basic
physics of strongly interacting matter related to a given
energy domain. The central horizontal bar indicates the
energy range of the CERN SPS.
}%
\label{ion_physics}
\end{figure}

High energy nucleus-nucleus collisions have been studied
experimentally since the beginning of 1970s in several
international and national laboratories.
Present and future accelerator facilities for
relativistic nuclear beams are summarized in Fig.~\ref{landscape},
where their nominal energy range is also given.
They are ordered according to the top collision energy
in Fig.~\ref{ion_physics}. The basic physics of the strongly
interacting matter which is related to  a given energy domain
is also given in Fig.~\ref{ion_physics}.

In addition to the CERN SPS
the physics of the onset of deconfinement and the critical point
can be studied at the BNL RHIC \cite{rhic_low}.

Up to now the experiments at RHIC were performed in the
collider mode in the energy range $\sqrt{s_{NN}} = 20 - 200$ GeV.
It seems, however, plausible to run RHIC even at significantly lower energies \cite{todd},
down to  $\sqrt{s_{NN}} \approx 5$ GeV, and thus cover the SPS energy range.
The advantages of this potential RHIC program would be:
\begin{itemize}
\item
a broad collision energy range covered by a single experimental facility
which yields  small relative systematic errors of the resulting energy
dependence of hadron production properties,
\item
a uniform and almost complete acceptance around midrapidity,
provided
the existing STAR apparatus is used in this study; it is important
for the measurements of the anisotropic flow.
\end{itemize}
However, running experiments in the collider mode does not
allow a measurement of the spectator fragments and thus a
selection of events with a fixed number of interacting nucleons.
The latter is important for event-by-event fluctuation studies,
in particular fluctuations of extensive variables 
like particle multiplicity \cite{soft_balls}.

The RHIC accelerator can be used also as a fixed target machine.
In this case the top energy is $\sqrt{s_{NN}} \approx 14$ GeV
and the existing BRAHMS detector can be used for measurements of
inclusive identified hadron spectra \cite{rhic_low}.
This option is considered as a possible fast cross-check of the
NA49 results.

In view of
the importance of the physics of the onset of deconfinement
and the critical point it appears desirable to perform
the experiments in both centers. The expected results will be
complementary in the important aspects of fluctuations and
anisotropic flow. The partial overlap of the results will allow
the necessary experimental cross-checks.

\section{Summary}
The recent experimental and theoretical findings
strongly suggest that a further study of nuclear collisions
in the CERN SPS energy range is of particular importance.
The new measurements can answer the questions concerning the nature
of the onset of deconfinement and the existence and location
of the critical point.

Consequently,
a new experimental program to study hadron production in collisions
of nuclei at the CERN SPS is proposed. It will focus on measurements
of fluctuations and long correlations in
C+C, Si+Si and In+In interactions at 10$A$, 20$A$, 30$A$,
40$A$, 80$A$ and 158$A$ GeV
with the aim to identify properties
of the onset of deconfinement and search for the critical point of
strongly interacting matter.
A parallel complementary effort at the BNL RHIC is strongly 
recommended.

\vspace{0.5cm}
{\bf Acknowledgements}

This work is based on the Letter of Intent prepared by the
NA49-future Collaboration.
It was partly supported by Virtual Institute of Strongly
Interacting Matter (VI-146) of Helmholtz Association, Germany.

{}


\begin{thebibliography}{}
\bibitem{cern50}
  R. Aymar et al.,
  Phys.\ Rept.\  {\bf 403-404}, 1 (2004).

\bibitem{villars}
  J. Dainton et al., [The CERN SPS and PS Committee],
  CERN--SPSC--2005--010, SPSC--M--730  (2005).

\bibitem{afanasiev:2002mx}
  S.~V.~Afanasiev {\it et al.}  [The NA49 Collaboration],
  Phys.\ Rev.\ C {\bf 66}, 054902 (2002)
  [arxiv:nucl-ex/0205002].

\bibitem{Gazdzicki:2004ef}
  M.~Gazdzicki {\it et al.}  [NA49 Collaboration],
  J.\ Phys.\ G {\bf 30}, S701 (2004)
  [arXiv:nucl-ex/0403023].

\bibitem{Gazdzicki:1998vd}
  M.~Gazdzicki and M.~I.~Gorenstein,
  Acta Phys.\ Polon.\ B {\bf 30}, 2705 (1999)
  [arXiv:hep-ph/9803462].

\bibitem{eoi}
J. Bartke et al.,
{\it A new experimental programme with nuclei and proton beams
at the CERN SPS}, CERN-SPSC-2003-038(SPSC-EOI-01) and presentations
at the Villars workshop 2004;
N. Antoniou et al., 
{\it Study of hadron production in collisions of protons and
nuclei at the CERN SPS}, the Letter of Intent in preparation. 

\bibitem{na49_nim}
  S.~Afanasev {\it et al.}  [NA49 Collaboration],
  Nucl.\ Instrum.\ Meth.\ A {\bf 430}, 210 (1999).

\bibitem{na49_beam}
  C.~Alt {\it et al.}  [NA49 Collaboration],
  Phys.\ Rev.\ Lett.\  {\bf 94}, 052301 (2005)
  [arXiv:nucl-ex/0406031] and
  P.~Dinkelaker  [NA49 Collaboration],
  J.\ Phys.\ G {\bf 31}, S1131 (2005).

\bibitem{Alt:2005zq}
  C.~Alt,
  arXiv:hep-ex/0510009.

\bibitem{qgp_sps}
  U.~W.~Heinz and M.~Jacob,
  arXiv:nucl-th/0002042,
  J.~Rafelski and B.~Muller,
  Phys.\ Rev.\ Lett.\  {\bf 48}, 1066 (1982)
  [Erratum-ibid.\  {\bf 56}, 2334 (1986)],
  T.~Matsui and H.~Satz,
  Phys.\ Lett.\ B {\bf 178}, 416 (1986),
  F.~Becattini, L.~Maiani, F.~Piccinini, A.~D.~Polosa and V.~Riquer,
  arXiv:hep-ph/0508188.
\bibitem{qgp_rhic}
  I.~Arsene {\it et al.}  [BRAHMS Collaboration],
  Nucl.\ Phys.\ A {\bf 757}, 1 (2005)
  [arXiv:nucl-ex/0410020],
  B.~B.~Back {\it et al.},
  Nucl.\ Phys.\ A {\bf 757}, 28 (2005)
  [arXiv:nucl-ex/0410022],
  J.~Adams {\it et al.}  [STAR Collaboration],
  Nucl.\ Phys.\ A {\bf 757}, 102 (2005)
  [arXiv:nucl-ex/0501009],
  K.~Adcox {\it et al.}  [PHENIX Collaboration],
  Nucl.\ Phys.\ A {\bf 757}, 184 (2005)
  [arXiv:nucl-ex/0410003],

  \bibitem{Gorenstein:2003cu}
  M.~I.~Gorenstein, M.~Gazdzicki and K.~A.~Bugaev,
  Phys.\ Lett.\ B {\bf 567}, 175 (2003)
  [arXiv:hep-ph/0303041].

\bibitem{hama}
  Y.~Hama, F.~Grassi, O.~Socolowski, T.~Kodama, M.~Gazdzicki and M.~Gorenstein,
  Acta Phys.\ Polon.\ B {\bf 35}, 179 (2004).

\bibitem{hadronic_models}
  E.~L.~Bratkovskaya {\it et al.},
  Phys.\ Rev.\ C {\bf 69}, 054907 (2004)
  [arXiv:nucl-th/0402026],
  J.~Cleymans and K.~Redlich,
  Phys.\ Rev.\ C {\bf 60}, 054908 (1999)
  [arXiv:nucl-th/9903063].
\bibitem{mg_fluct}
  M.~Gazdzicki, M.~I.~Gorenstein and S.~Mrowczynski,
  Phys.\ Lett.\ B {\bf 585}, 115 (2004)
  [arXiv:hep-ph/0304052] and
  M.~I.~Gorenstein, M.~Gazdzicki and O.~S.~Zozulya,
  Phys.\ Lett.\ B {\bf 585}, 237 (2004)
  [arXiv:hep-ph/0309142].

\bibitem{Kolb:2000sd}
  P.~F.~Kolb, J.~Sollfrank and U.~W.~Heinz,
  Phys.\ Rev.\ C {\bf 62}, 054909 (2000)
  [arXiv:hep-ph/0006129].


\bibitem{Cleymans:1999cb}
  J.~Cleymans and K.~Redlich,
  Nucl.\ Phys.\ A {\bf 661}, 379 (1999)
  [arXiv:nucl-th/9906065].
\bibitem{Braun-Munzinger:2003zd}
  P.~Braun-Munzinger, K.~Redlich and J.~Stachel,
  arXiv:nucl-th/0304013.

\bibitem{becat}
  F.~Becattini and U.~W.~Heinz,
  Z.\ Phys.\ C {\bf 76}, 269 (1997)
  [Erratum-ibid.\ C {\bf 76}, 578 (1997)]
  [arXiv:hep-ph/9702274] and
  F.~Becattini, M.~Gazdzicki and J.~Sollfrank,
  Eur.\ Phys.\ J.\ C {\bf 5}, 143 (1998)
  [arXiv:hep-ph/9710529].

  \bibitem{Rajagopal:2000wf}
  K.~Rajagopal and F.~Wilczek,
  arXiv:hep-ph/0011333.

\bibitem{Stephanov:2004wx}
  M.~A.~Stephanov,
  arXiv:hep-ph/0402115.

\bibitem{Karsch:2004wd}
  F.~Karsch,
  J.\ Phys.\ G {\bf 31}, S633 (2005)
  [arXiv:hep-lat/0412038].

\bibitem{katz}
  S.~D.~Katz,
  arXiv:hep-ph/0511166.

\bibitem{jakko}
  F.~Becattini, J.~Manninen and M.~Gazdzicki,
  arXiv:hep-ph/0511092.

\bibitem{Gorenstein:2005rc}
  M.~I.~Gorenstein, M.~Gazdzicki and W.~Greiner,
  Phys.\ Rev.\ C {\bf 72}, 024909 (2005)
  [arXiv:nucl-th/0505050].

\bibitem{antoniou:2005}
N. G. Antoniou, Y. F. Contoyiannis, F. K. Diakonos, A. I. Karanikas,
C. N. Ktorides, Nucl. Phys. A {\bf 693}, 799 (2001),
N. G. Antoniou,  F. K. Diakonos, G. Mavromanolakis,
Nucl. Phys. A {\bf 761}, 149 (2005)

\bibitem{Hatta:2002sj}
  Y.~Hatta and T.~Ikeda,
  Phys.\ Rev.\ D {\bf 67}, 014028 (2003)
  [arXiv:hep-ph/0210284].

\bibitem{Fodor:2004nz}
  Z.~Fodor and S.~D.~Katz,
  JHEP {\bf 0404}, 050 (2004)
  [arXiv:hep-lat/0402006].

\bibitem{Allton:2005gk}
  C.~R.~Allton {\it et al.},
  Phys.\ Rev.\ D {\bf 71}, 054508 (2005)
  [arXiv:hep-lat/0501030].

\bibitem{Stephanov:1999zu}
  M.~A.~Stephanov, K.~Rajagopal and E.~V.~Shuryak,
  Phys.\ Rev.\ D {\bf 60}, 114028 (1999)
  [arXiv:hep-ph/9903292].

\bibitem{na49_flow}
  C.~Alt {\it et al.}  [NA49 Collaboration],
  Phys.\ Rev.\ C {\bf 68}, 034903 (2003)
  [arXiv:nucl-ex/0303001].

\bibitem{na49_pt}
  T.~Anticic {\it et al.}  [NA49 Collaboration],
  Phys.\ Rev.\ C {\bf 70}, 034902 (2004)
  [arXiv:hep-ex/0311009].

\bibitem{na49_n}
  M.~Rybczynski {\it et al.}  [NA49 Collaboration],
  J.\ Phys.\ Conf.\ Ser.\  {\bf 5}, 74 (2005)
  [arXiv:nucl-ex/0409009].

\bibitem{soft_balls}
  M.~Gazdzicki and M.~Gorenstein,
  arXiv:hep-ph/0511058.

\bibitem{na49_k}
  C.~Roland {\it et al.}  [NA49 Collaboration],
  J.\ Phys.\ G {\bf 30}, S1381 (2004)
  [arXiv:nucl-ex/0403035].

\bibitem{rhic_low}
  G.~Stephans, Proceedings of the Quark Matter 2005, August  2005,
  Budapest, Hungary.

\bibitem{todd}
  T.~Satogata, private communication.

\end{thebibliography}
\end{document}